\documentclass[modern]{aastex62}
\usepackage{threeparttable}

\newcommand{\ktwo}{\textit{K2}}
\newcommand{\kms}{km~s$^{-1}$}
\newcommand{\ms}{m~s$^{-1}$}

\newcommand{\masyr}{mas~yr$^{-1}$}
\newcommand{\err}{\textit{$\pm$}}
\newcommand{\teff}{$T_\mathrm{eff}$}
\newcommand{\msun}{$M_\odot$}
\newcommand{\rsun}{$R_\odot$}
\newcommand{\lsun}{$L_\odot$}

\newcommand{\rearth}{$R_\oplus$}

\newcommand{\thestar}{K2-233}

\newcommand{\gaia}{\textit{Gaia}}
\newcommand{\kepler}{\textit{Kepler}}

\newcommand{\tess}{\textit{TESS}}

\received{February 2, 2018}
\revised{March 13, 2018}
\accepted{April 10, 2018}
\submitjournal{AAS Journals}

\shorttitle{Three small planets around \thestar}
\shortauthors{David et al.}

\begin{document}

\title{Three small planets transiting the bright young field star \thestar}

\correspondingauthor{Trevor J.\ David}
\email{trevor.j.david@jpl.nasa.gov}

\author[0000-0001-6534-6246]{Trevor J.\ David}
\affil{Jet Propulsion Laboratory, California Institute of Technology, 4800 Oak Grove Drive, Pasadena, CA 91109, USA}

\author[0000-0002-1835-1891]{Ian J.~M.\ Crossfield}
\affil{Department of Physics, Massachusetts Institute of Technology, Cambridge, MA, USA}

\author[0000-0001-5578-1498]{Bj\"{o}rn Benneke} 
\affil{D\'{e}partement de Physique, Universit\'{e} de Montr\'{e}al, Montreal, H3T J4, Canada}

\author[0000-0003-0967-2893]{Erik A.\ Petigura}
\altaffiliation{NASA Hubble Fellow}
\affil{Department of Astronomy, California Institute of Technology, Pasadena, CA 91125, USA}

\author{Erica J.\ Gonzales}
\altaffiliation{NSF Graduate Research Fellow}
\affil{Astronomy and Astrophysics Department, University of California, Santa Cruz, CA, USA}

\author[0000-0001-5347-7062]{Joshua E.\ Schlieder}
\affil{Exoplanets and Stellar Astrophysics Laboratory, Code 667, NASA Goddard Space Flight Center, Greenbelt, MD 20771, USA}

\author{Liang\ Yu}
\affil{Department of Physics, Massachusetts Institute of Technology, Cambridge, MA, USA}

\author[0000-0002-0531-1073]{Howard T.\ Isaacson}
\affil{Astronomy Department, University of California, Berkeley, CA 94720, USA}

\author[0000-0001-8638-0320]{Andrew W.\ Howard}
\affil{Department of Astronomy, California Institute of Technology, Pasadena, CA 91125, USA}

\author[0000-0002-5741-3047]{David R.\ Ciardi}
\affil{Caltech/IPAC-NASA Exoplanet Science Institute, Pasadena, CA 91125, USA}

\author[0000-0003-2008-1488]{Eric E.\ Mamajek}
\affil{Jet Propulsion Laboratory, California Institute of Technology, 4800 Oak Grove Drive, Pasadena, CA 91109, USA}
\affil{Department of Physics \& Astronomy, University of Rochester, Rochester, NY 14627, USA}

\author{Lynne A.\ Hillenbrand}
\affil{Department of Astronomy, California Institute of Technology, Pasadena, CA 91125, USA}

\author[0000-0002-3656-6706]{Ann Marie Cody}
\affil{NASA Ames Research Center, Moffet Field, CA 94035, USA}

\author[0000-0003-1645-8596]{Adric Riedel}
\affil{Space Telescope Science Institute, 3700 San Martin Drive, Baltimore, MD 21218, USA}

\author{Hans Martin Schwengeler}
\affil{Planet Hunter, Bottmingen, Switzerland}

\author{Christopher Tanner}
\affil{Exoplanet Explorers Volunteer}

\author{Martin Ende}
\affil{Exoplanet Explorers Volunteer}

\begin{abstract}
We report the detection of three small transiting planets around the young K3 dwarf K2-233 (2MASS J15215519-2013539) from observations during Campaign 15 of the \ktwo\ mission. The star is relatively nearby ($d$ = 69~pc) and bright ($V = 10.7$~mag, $K_s = 8.4$~mag), making the planetary system an attractive target for radial velocity follow-up and atmospheric characterization with the \textit{James Webb Space Telescope}. The inner two planets are hot super-Earths ($R_b$ = 1.40 $\pm$ 0.06~\rearth, $R_c$ = 1.34 $\pm$ 0.08~\rearth), while the outer planet is a warm sub-Neptune ($R_d$ = 2.6 $\pm$ 0.1~\rearth). We estimate the stellar age to be 360$^{+490}_{-140}$ Myr based on rotation, activity, and kinematic indicators. The K2-233 system is particularly interesting given recent evidence for inflated radii in planets around similarly-aged stars, a trend potentially related to photo-evaporation, core-cooling, or both mechanisms.
\end{abstract}


\section{Introduction} \label{sec:intro}
\kepler\ provided a large, relatively homogeneous sample from which the statistical frequencies of exoplanets have been robustly determined. Though the primary mission was to measure the prevalence of Earth-sized planets around solar-type stars \citep[e.g.][]{Petigura:etal:2013b}, the data also provided a number of insights into planet formation outcomes more generally, such as the surprising abundance of sub-Neptunes, trends in planet occurrence with stellar mass \citep{Howard:etal:2012}, and fine structure in the size distribution of small planets \citep{Fulton:etal:2017}. Because the mission surveyed $\sim$1/400 of the sky, however, a typical \kepler\ planet host is relatively faint and presents challenges to characterization efforts like radial velocity mass measurements or transmission spectroscopy.

In contrast, the \ktwo\ mission \citep{Howell:etal:2014} to date has observed 15$\times$ the area of the prime \kepler\ mission, casting a wider net for planets around bright stars more evenly distributed on the sky. Statistical exoplanet studies within carefully defined sub-samples may yet prove fruitful but, like its predecessor, \ktwo\ has already revealed a great number of surprises: a transiting minor planet around a stellar remnant \citep{Vanderburg:etal:2015}, the possible detection of accretion pulses driven by the orbital motion of an infant hot Jupiter \citep{Biddle:etal:2018}, and a chain of five near-resonant planets discovered by citizen scientists \citep{Christiansen:2018}, to name a few.

Two domains probed by \ktwo, which will form an important part of the mission's legacy, are transiting planet hosts that are bright and/or young. The mission has yielded all of the known transiting planets in young clusters and associations to date \citep[see][for a review]{Rizzuto:etal:2017}, as well as a number of planets around active field stars that are likely to be moderately young \citep[e.g.][]{Gaidos:etal:2017, Dai:etal:2017, Barragan:etal:2018}. \ktwo\ is also responsible for contributing some of the brightest known transiting planet hosts, such as HIP 41378 \citep{Vanderburg:etal:2016b}, HD 106315 \citep{Crossfield:etal:2017, Rodriguez:etal:2017}, HD 3167 \citep{Vanderburg:etal:2016c, Gandolfi:etal:2017}, and GJ 9827 \citep{Rodriguez:etal:2018, Niraula:etal:2017}. Until \tess\ extends the sample of bright transiting planet hosts, these systems remain some of the most amenable to atmospheric characterization via transit transmission spectroscopy. The properties of young planets are particularly interesting, given suggestions that larger sub-Neptunes are preferentially found around young stars \citep{Berger:etal:2018} and that such planets may experience evaporative mass-loss at early stages \citep[e.g.][]{Lopez:Fortney:2013}. Here we report the discovery of three small transiting planets around a star that is both relatively bright and young. The system, \thestar, is an attractive target for both radial velocity work and atmospheric characterization.

\section{\ktwo\ Observations} \label{sec:observations}
K2-233 (EPIC 249622103, 2MASS J15215519-2013539) was observed during Campaign 15 of the \ktwo\ mission\footnote{The star was proposed by several K2 teams:  GO15020 (PI Adams), GO15023 (PI Hillenbrand), GO15043 (PI Rizutto), and GO15052 (PI Stello).}. Following an approach similar to that of \cite{Christiansen:2018}, we analyzed the raw cadence pixel data released by the \ktwo\ project by first converting the cadence data into target pixel files with \texttt{kadenza}\footnote{\url{https://github.com/KeplerGO/kadenza}} \citep{geert_barentsen_2017_344973}.  From there we followed our team's standard discovery approach \citep[see e.g.\ ][]{Crossfield:2016}: we constructed a light curve from aperture photometry with \texttt{k2phot},\footnote{\url{https://github.com/petigura/k2phot}} which simultaneously models stellar variability and spacecraft systematics with a Gaussian process. From this light curve we found three transit signals detected with the \texttt{terra}\footnote{\url{https://github.com/petigura/terra}} program \citep{Petigura:etal:2013b,Petigura:etal:2013a}. The transit signals were also independently discovered by citizen scientists as part of the Planet Hunters and Exoplanet Explorers projects.\footnote{\url{https://www.planethunters.org/}}\footnote{\url{https://www.zooniverse.org/projects/ianc2/exoplanet-explorers}} 

Prior to fitting transit models to the \ktwo\ data, we removed the stellar variability via cubic basis spline fits with iterative outlier rejection. We used the RMS in the flattened light curve as the flux uncertainty for each measurement. We then fit analytic transit models, generated with the \texttt{PyTransit}\footnote{\url{https://github.com/hpparvi/PyTransit}} package \citep{Parviainen:2015}, to the observations in order to determine the following free parameters: the orbital period $P$, time of mid-transit $T_0$, radius ratio $R_P/R_*$, scaled semi-major axis $a/R_*$, and cosine of the inclination $\cos{i}$. We first performed Levenberg-Marquardt (L-M) fits to find initial parameter estimates, then used the \texttt{emcee} affine invariant implementation of the Markov chain Monte Carlo (MCMC) method \citep{Foreman-Mackey:etal:2013} to robustly determine the uncertainties on these parameters. The target probability density to be sampled in these simulations was:

\begin{equation}
    \ln{\mathcal{L}} = -\frac{1}{2} \chi^2 - \frac{1}{2}\frac{(\rho_*-\mu_{\rho_{*}})^2}{\sigma_{\rho_{*}}^2},
\end{equation}

where the first term is the likelihood and the second term describes a Gaussian prior on the mean stellar density, $\rho_*$, with $\mu_{\rho_{*}}$ = 2.73~g cm$^{-3}$ and $\sigma_{\rho_{*}}$ = 0.31~g cm$^{-3}$, based on our stellar characterization in \S~\ref{sec:star}. We used $\chi^2 = \sum (f_n-m_n)^2/\sigma_n^2$, where $f_n$ and $m_n$ are the $n$-th flux observation and transit model values, respectively, and $\sigma_n$ is the individual flux uncertainty. This assumes uncorrelated measurement uncertainties, which is not strictly true due to e.g. short-term stellar variability and our procedure of removing the stellar variability prior to fitting. Additionally, we imposed uniform priors on the following parameters: $P$ (centered on the initial estimate, with width 0.01 d), $T_0$ (centered on the initial estimate with width 0.06$P$), $R_P/R_*$ (from -1 to +1), $a/R_*$ (from 1 to $\infty$), $\cos{i}$ (from $\cos{50^\circ}$ to $\cos{90^\circ}$). We assumed a quadratic limb darkening law with coefficients $u_1$=0.587 and $u_2$=0.136, informed by our spectroscopic stellar characterization (\S~\ref{sec:star}) and the values tabulated by \citet{Claret:Bloemen:2011}. The transit models were numerically integrated to match the \kepler\ long cadence integration using a super-sampling factor of 10. 

We initialized the MCMC sampler with 40 walkers around the preliminary L-M solution. For each free parameter the integrated autocorrelation time, $\hat{\tau}$, of the MCMC chain was calculated every 2000 steps. When the chain length exceeded 100$\times \hat{\tau}$ for all parameters and when these $\hat{\tau}$ estimates changed by less than 1\% the chain was considered to be converged and the MCMC procedure was halted. We estimated the burn-in as 5 times the maximum autocorrelation time estimate, and from the trimmed chains we calculated the 15.87, 50, 84.13 percentile values for each parameter.\footnote{Transit fit posteriors are available at \url{https://exofop.ipac.caltech.edu/k2/edit_target.php?id=249622103}.} These parameters and derivative physical quantities are reported in Table~\ref{table:planet}.

We additionally performed fits with the eccentricity $e$ and periastron longitude $\omega$ as free parameters with uniform priors on each. As expected, the relatively shallow transits provide only weak constraints on eccentricity of $e <$ 0.53, 0.54, 0.45 at 95\% confidence for planets b, c, and d respectively. Orbit crossing constraints would restrict the range of allowed eccentricities to even smaller values. We ultimately adopted the circular model because it is simpler (lower BIC), the other fitted parameters changed by $<1\sigma$, and previous studies of compact multiplanet systems find typical eccentricities of a few percent \citep{Hadden:Lithwick:2014, vanEylen:Albrecht:2015, Xie:etal:2016}. We also investigated the effect of eliminating the Gaussian prior on the mean stellar density prior (requiring only that $a/R_*>$ 1). These fits, which are presented in the appendix, resulted in $R_P/R_*$ distributions with longer tails towards more positive values, corresponding to solutions with higher impact parameters and unrealistically low stellar densities. The median parameter values from these fits all changed by $\lesssim 1\sigma$ and the stellar density implied by each planet was within $1\sigma$ of the value we adopted, providing assurance that the star has been properly characterized and our prior on this parameter is well-justified.
 
\begin{figure}
    \centering
    \includegraphics[width=\linewidth]{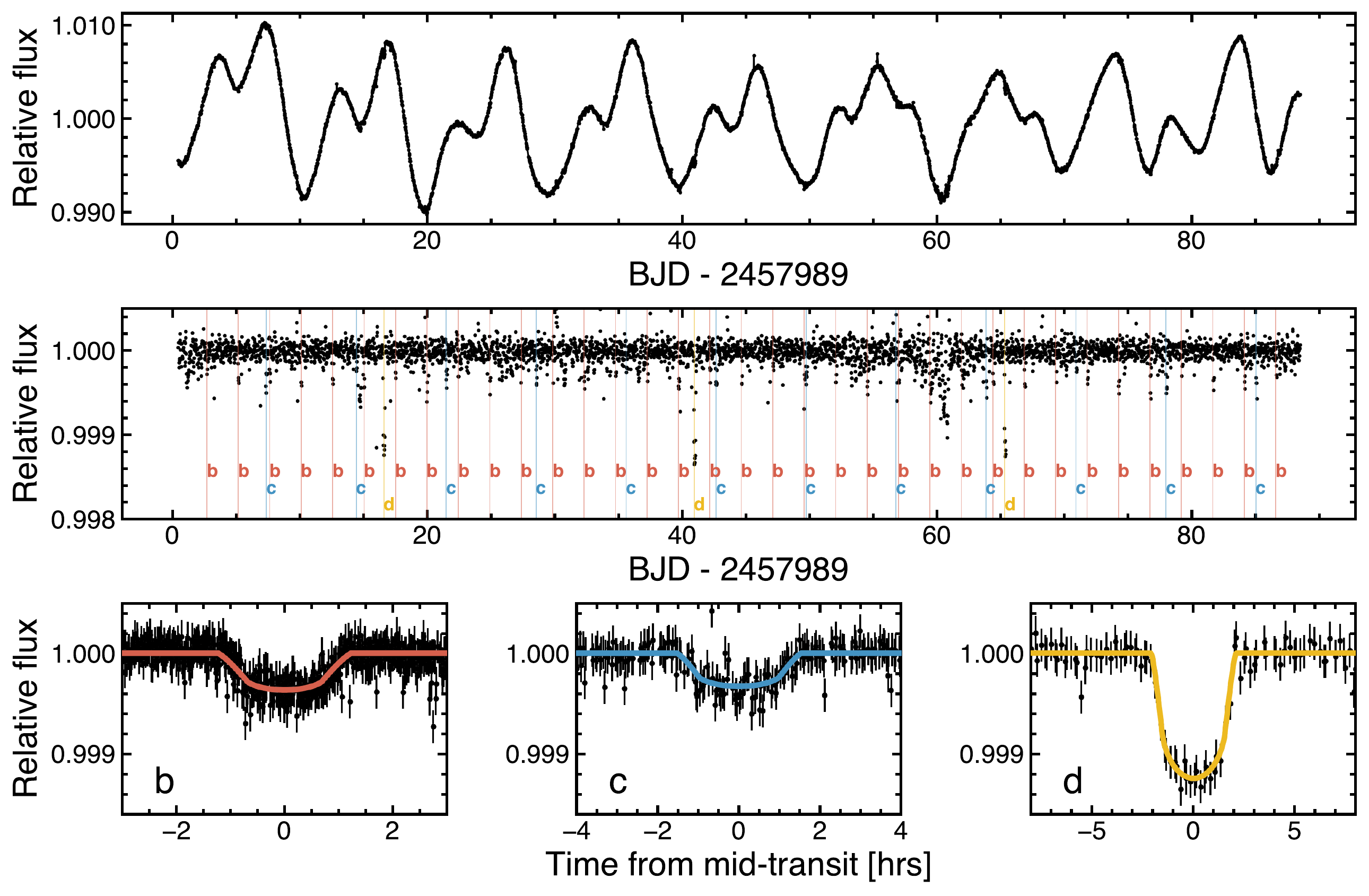}
    \caption{Full \ktwo\ light curve of \thestar\ (top), with the stellar variability removed and individual transits shown (middle), and phase-folded to the transits of planets b, c, and d with transit model fits shown as shaded lines (bottom).}
    \label{fig:lc}
\end{figure}

\begin{deluxetable*}{lccc}
\tablecaption{Planet parameters in the \thestar\ system \label{table:planet}}
\tablecolumns{4}
\tablewidth{-0pt}
\tabletypesize{\footnotesize}
\tablehead{
        \colhead{Parameter} &
        \colhead{Planet b} & 
        \colhead{Planet c} & 
        \colhead{Planet d} \\
        }
\startdata
\textit{Directly fitted parameters} \\
Time of mid-transit, $T_0$ (BJD-2450000) & 7991.6910$^{+0.0026}_{-0.0025}$ & 7996.3522$^{+0.0056}_{-0.0057}$ & 8005.5801$^{+0.0025}_{-0.0024}$ \\
Orbital period, $P$ (days) & 2.46746$^{+0.00014}_{-0.00014}$ & 7.06142$^{+0.00084}_{-0.00084}$ & 24.3662$^{+0.0021}_{-0.0021}$ \\
Radius ratio, $R_P/R_*$                      & 0.01721$^{+0.00049}_{-0.00047}$ & 0.01643$^{+0.00084}_{-0.00078}$ & 0.03254$^{+0.00080}_{-0.00079}$ \\
Scaled semi-major axis, $a/R_*$              &  9.49$^{+0.29}_{-0.32}$ & 19.34$^{+0.67}_{-0.74}$ & 44.2$^{+1.6}_{-1.8}$ \\
Cosine of inclination, $\cos{i}$ & 0.021$^{+0.016}_{-0.014}$ & 0.0184$^{+0.0082}_{-0.0107}$ & 0.0113$^{+0.0019}_{-0.0021}$ \\
\\
\textit{Derived parameters} \\
Planet radius, $R_P$ (R$_\oplus$) & 1.398$^{+0.062}_{-0.060}$ & 1.335$^{+0.083}_{-0.077}$ & 2.64$^{+0.11}_{-0.11}$ \\
Inclination, $i$ (deg) & 88.79$^{+0.82}_{-0.94}$ & 88.95$^{+0.61}_{-0.47}$ & 89.35$^{+0.12}_{-0.11}$ \\
Impact parameter, $b$ & 0.20$^{+0.15}_{-0.14}$ & 0.36$^{+0.15}_{-0.21}$ & 0.500$^{+0.065}_{-0.079}$ \\
Total duration, $T_{14}$ (hrs) & 1.969$^{+0.062}_{-0.063}$ & 2.65$^{+0.13}_{-0.15}$ & 3.808$^{+0.089}_{-0.088}$ \\
Full duration, $T_{23}$ (hrs) & 1.899$^{+0.062}_{-0.065}$ & 2.55$^{+0.14}_{-0.17}$ & 3.49$^{+0.10}_{-0.10}$ \\
Semi-major axis, $a$ (AU) & 0.03317$^{+0.00044}_{-0.00045}$ & 0.06687$^{+0.00088}_{-0.00090}$ & 0.1527$^{+0.0020}_{-0.0021}$ \\
Insolation flux, $S$ (S$_\oplus$) & 273$^{+30}_{-30}$ & 67.1$^{+7.4}_{-7.3}$ & 12.9$^{+1.4}_{-1.4}$ \\
Equilibrium temperature, $T_\mathrm{eq}$ (K)$^{a}$ & 1040$^{+28}_{-26}$ & 728$^{+20}_{-19}$ & 482$^{+14}_{-13}$ \\
\enddata
\begin{tablenotes}
\item Reported values and 1$\sigma$ errors are the 50, 15.87, and 84.13 percentile levels from the MCMC chain. The fit presented here assumed $e=0$, a Gaussian prior on the mean stellar density ($\mu_{\rho_{*}}$ = 2.73~g cm$^{-3}$, $\sigma_{\rho_{*}}$ = 0.31~g cm$^{-3}$), quadratic limb darkening coefficients $u_1$ = 0.587, $u_2$ = 0.136, and no contaminating flux. Derived parameters assume all three planets are orbiting the target star, and that the target star is single.
\item (a) Assuming an albedo of 0.3. 
\end{tablenotes}
\end{deluxetable*}

\section{Stellar characterization} \label{sec:star}

We acquired high-resolution optical spectroscopy of \thestar\ on UT 2018 Jan 22 (BJD 2458141.152490) with Keck-I/HIRES \citep{Vogt:etal:1994} using standard procedures of the California Planet Search \citep{Howard:etal:2010}. We then used SpecMatch \citep{Petigura:2015} to compare the spectrum to \citet{Coelho:etal:2005} model atmospheres and determined \teff\  = 4950 $\pm$ 100~K, $\log{g}$ = 4.71 $\pm$ 0.10~dex, [Fe/H] = 0.07 $\pm$ 0.06~dex, and $v\sin{i_*}$ = 4.5 $\pm$ 1.0~\kms. As a consistency check, we also used SpecMatch-Emp \citep{Yee:etal:2017} to compare the spectrum with a library of empirical spectra of benchmark stars, finding values for \teff, $R_*$, and [Fe/H] that are consistent within 1$\sigma$ of those found with SpecMatch. From this analysis we found the best-matching template spectrum to be that of HD 110463 (K3V), from which we assigned a spectral type of K3 to \thestar. A precise distance to \thestar\ has been measured from trigonometric parallax (69 $\pm$ 1~pc; \gaia\ DR1), which provides tight constraints on the stellar parameters. With the \texttt{isoclassify} package \citep{Huber:etal:2017}, using the parallax, spectroscopic parameters (\teff, $\log{g}$, and [Fe/H]), and the 2MASS $JHK_s$ magnitudes as input, we determined precise values for the model-dependent mass, $M_*$ = 0.80 $\pm$ 0.02~\msun, and radius, $R_*$ = 0.745 $\pm$ 0.011~\rsun. To account for possible systematic uncertainties in the models, we added in quadrature a 2\% uncertainty in these parameters. Combined with our spectroscopic \teff, the stellar radius and Stefan-Boltzmann law imply a luminosity of $L_*$ = 0.300 $\pm$ 0.032~\lsun. 

In an effort to better constrain the age of the system, we next considered the stellar kinematics, rotation, activity, and spectroscopic age indicators. The barycentric radial velocity was measured to be $-$9.73 $\pm$ 0.20~\kms\ using the telluric A and B absorption bands as a wavelength reference \citep{Chubak:etal:2012}. Combined with the proper motions and distance from \gaia, we found the stellar kinematics are not a good match to any known moving groups, nearby open clusters, or star-forming regions. Using the BANYAN $\Sigma$ tool \citep{Gagne:etal:2018}, we found a 68.5\% probability that the star belongs to the field population, with the remaining 31.5\% probability assigned to membership in the Upper Scorpius OB association. The mean distance to Upper Scorpius is $\sim$140~pc, with a unidirectional spread of $\lesssim$35~pc. Given the precisely determined distance (at half the mean distance to Upper Scorpius), a radius and mean stellar density consistent with a main sequence star, and lack of lithium absorption (discussed below), we can confidently rule out that \thestar\ belongs to that association. Based on age diagnostics from the stellar rotation and activity we suggest \thestar\ is a young field star. 

\begin{figure}
    \centering
    \includegraphics[width=\linewidth]{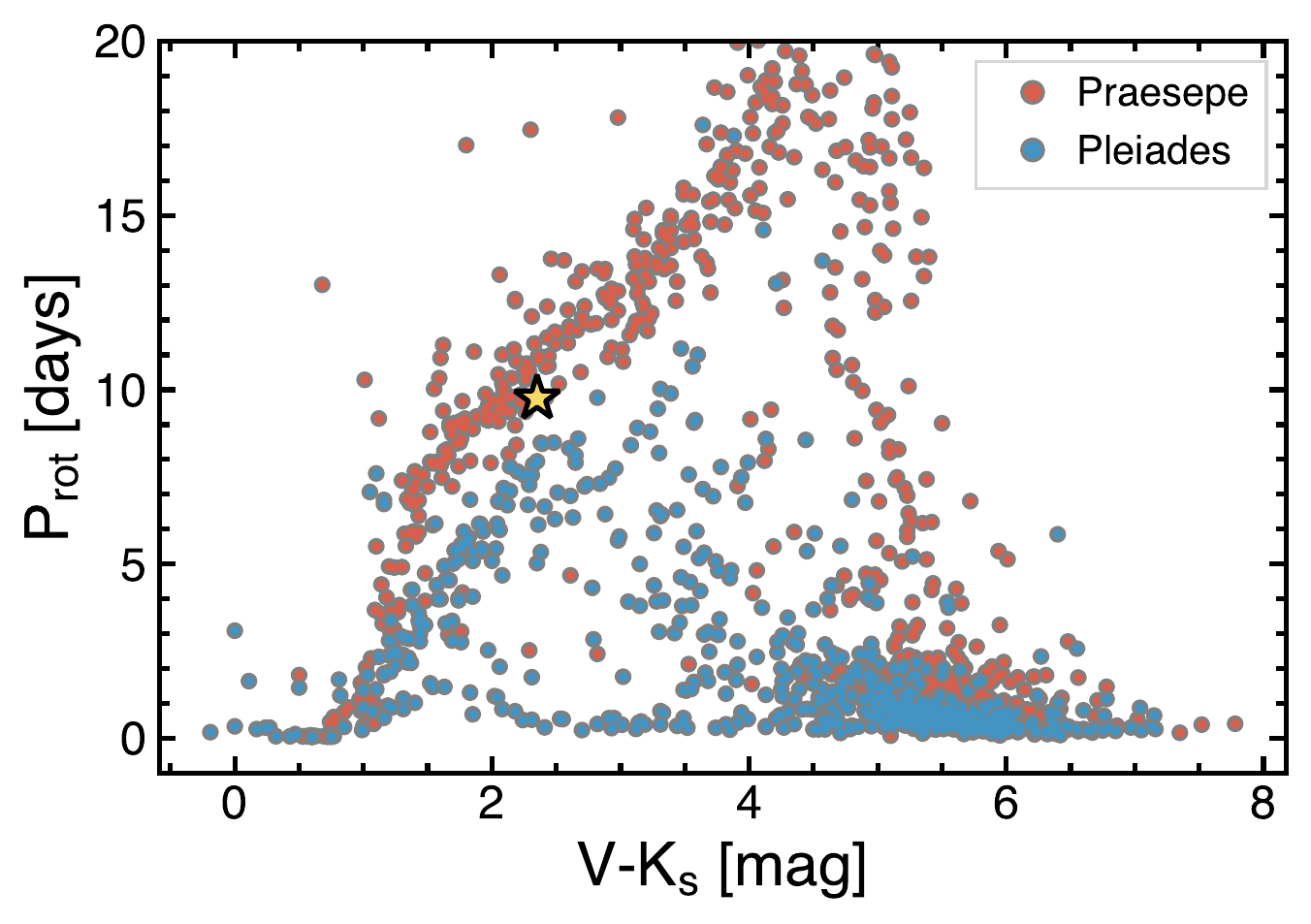}
    \caption{Period-color diagram for open cluster members observed by \ktwo\ \citep{Rebull:etal:2016,Rebull:etal:2017} and \thestar\ (gold star).}
    \label{fig:prot}
\end{figure}

From an autocorrelation function (ACF) analysis of the light curve, we measured a rotation period of $P_\mathrm{rot} = 9.754 \pm 0.038$~d. The period was determined by the slope of a linear fit to the first four peaks of the ACF plus the origin, and the uncertainty was estimated from the scatter about that fit. We note that typical observed rates of surface differential rotation ($\lesssim 0.07$~rad~d$^{-1}$) in dwarf stars of a similar temperature and rotation period might introduce an additional uncertainty in the rotation period of $\lesssim$1.4~d. In period-color space (Figure~\ref{fig:prot}), the star is situated between members of the Pleiades ($\sim$125~Myr) and Praesepe ($\sim$790~Myr), suggesting an age intermediate to these clusters if the star is on the main sequence \citep{Rebull:etal:2016, Rebull:etal:2017}. The variability amplitude, 0.014 mag (measured from the 10th to the 90th percentile) is also similar to those seen amongst Pleiades and Praesepe members of a similar color \citep{Rebull:etal:2016, Rebull:etal:2017}. Different gyrochronology relations predict ages of 270$^{+80}_{-70}$~Myr \citep{Barnes:2007}, 440$^{+120}_{-110}$~Myr \citep{Mamajek:Hillenbrand:2008}, and 500$^{+140}_{-120}$~Myr \citep{Angus:etal:2015}, where these estimates reflect the 16th, 50th, and 84th percentile values adopting a conservative error on the rotation period of 1.4~d to allow for the possibility of differential rotation. Existing gyrochronology relations are in need of re-calibration, so these ages should be regarded with caution, but all relations suggest an age younger than 1~Gyr. 

The HIRES spectrum shows H$\alpha$ in absorption and no detectable absorption at \ion{Li}{1}~6708~\AA, which argues for an age older than that of the Pleiades. From the spectrum, we also measured the $S_{HK}$ index and $\log{R^{'}_{HK}}$ = -4.361~dex using the method of \citet{Isaacson:etal:2010}, which is a value typical for Hyades ($\sim$625--850~Myr) and the Ursa Major moving group ($\sim$400--600~Myr) members of a similar color \citep[see Figures 4 and 5 of][]{Mamajek:Hillenbrand:2008}. From the \citet{Mamajek:Hillenbrand:2008} activity-age relation we calculated an age of 220~Myr, consistent with the younger gyrochronology age quoted above. The age relations considered are here are statistical in nature and carry large uncertainties. New calibrations of age-activity and gyrochronology relations are clearly in order, but outside the scope of this paper. We ultimately adopt an age of $\tau_* \approx$ 360$^{+490}_{-140}$~Myr, corresponding to the mean of the four estimates above, with the lower bound originating from the activity age and the upper bound from the oldest ages suggested for the Hyades and Praesepe clusters. We summarize the basic observables and results of our stellar characterization analyses in Table~\ref{table:star}.

\begin{deluxetable*}{lcl}
\tablecaption{Parameters of K2-233 \label{table:star}}
\tablecolumns{3}
\tablewidth{-0pt}
\tabletypesize{\scriptsize}
\tablehead{
        \colhead{Parameter} &
        \colhead{Value} & 
        \colhead{Source} \\
        }
\startdata
\textit{Kinematics and position}\\
R.A., J2000 (hh mm ss) & 15 21 55.198 & A\\
Dec., J2000 (dd mm ss) & -20 13 53.991 & A\\
Parallax (mas) & 14.50 \err\ 0.23 & A\\
Distance (pc) & 69 \err\ 1 & A\\
$\mu_\alpha$ (\masyr) & -20.174 \err\ 0.687 & A\\
$\mu_\delta$ (\masyr) & -30.921 \err\ 0.412 & A\\
Barycentric RV (\kms) & -9.73 \err\ 0.20 & B \\
\\
\textit{Photometry}\\
$G$ (mag) & 10.333 \err\ 0.001 & A \\
$B$ (mag) & 11.664 \err\ 0.027 & C \\
$V$ (mag) & 10.726 \err\ 0.019 & C \\
$J$ (mag) & 8.968 \err\ 0.020 & D \\
$H$ (mag) & 8.501 \err\ 0.026 & D \\
$K_s$ (mag) & 8.375 \err\ 0.023 & D \\
\\
\textit{Physical properties}\\
Spectral type & K3 & B\\
$T_\mathrm{spec}$ (K) &  4950 \err\ 100 & B\\
$M_*$ (\msun) & 0.800 \err\ 0.032 & B\\
$R_*$ (\rsun) & 0.745 \err\ 0.025 & B\\
$L_*$ (\lsun) & 0.300 \err\ 0.032 & B\\
$\log{g}$ (dex) & 4.71 \err\ 0.10 & B\\
$\mathrm{[Fe/H]}$ (dex) & 0.07 \err\ 0.06 & B\\
$v\sin{i_*}$ (\kms) & 4.5 \err\ 1.0 & B\\
$P_\mathrm{rot}$ (d) & 9.754 \err\ 0.038 & B\\
$S_{HK}$ & 0.686 & B\\
$\log{R^{'}_{HK}}$ (dex) & -4.36 & B\\
$\tau_*$ (Myr) & 360$^{+490}_{-140}$ & B\\
\enddata
\begin{tablenotes}
\item[] A: \gaia\ DR1, B: this work, C: APASS DR9, D: 2MASS.
\end{tablenotes}
\end{deluxetable*}

\section{Validating the planets}
\label{sec:validation}
The \ktwo\ photometry were extracted from a rectangular aperture 24\arcsec $\times$ 36\arcsec\ in size. Pan-STARRS1 imaging shows there are no stars of comparable brightness within 1 arcminute, excluding the possibility that the transit signals arise from a widely separated companion. We acquired high spatial resolution imaging in the Br-$\gamma$ band with Keck-II/NIRC2 on UT 2017 Dec 29 and found no evidence for additional closely-projected sources. Our 5$\sigma$ contrast limits rule out sources with $\Delta$mag$<$4 outside of 0.15\arcsec\ and $\Delta$mag$<$8 from 1.7--3.8\arcsec. Using the \texttt{vespa} package \citep{Morton:2015} we statistically validated each planet, using 3$\times$ the light curve RMS as a conservative estimate of the maximum secondary eclipse depth and 0.1\arcsec\ (3$\times$ the NIRC2 resolution) as the photometric exclusion radius. From this analysis we found false positive probabilities of $1.9 \times 10^{-7}$, $9.6 \times 10^{-5}$, and $6.9 \times 10^{-7}$ for planets b, c, and d, respectively. Notably, these probabilities are calculated for each planet individually, and the overall false positive probability is in fact lower given the presence of multiple transiting planets.

We searched for secondary spectral lines in the HIRES spectrum using the method described in \citet{Kolbl:etal:2015}, and found no evidence for a projected companion within 0.8\arcsec\ down to 3\% the brightness of the primary star. We note this method is blind to companions with velocity separations $<$15~\kms. Further assurance that the transit signals originate from \thestar\ comes from the transit fits with no direct prior on the mean stellar density. For each planet, the median value for the stellar density was within 1$\sigma$ of the value we adopted for \thestar. While not conclusive, this observation is suggestive that the transiting planets are indeed orbiting \thestar. If \thestar\ is a binary that evaded our detection, the planetary radii might be larger by $\lesssim$20\%, given our vetting through high-resolution imaging and spectroscopy \citep{Ciardi:etal:2015}.

\section{Discussion} \label{sec:discussion}

Recent studies of \kepler\ multi-planet systems have found a high degree of intra-system uniformity \citep{Ciardi:etal:2013, Fabrycky:etal:2014, Milholland:etal:2017}. For example, planet sizes within an individual system are correlated, i.e. a planet is more likely to have a size similar to its neighbor than a size drawn at random from the observed distribution of planet sizes \citep[][, hereafter W18]{Weiss:etal:2018}. The W18 study also found that (1) in about 65\% of planet pairs in multi-transiting systems, the outer planet is larger than the inner planet, (2) planet separations are evenly spaced in log semi-major axis, and (3) adjacent planets tend to be separated by about 20 mutual Hill radii.

The \thestar\ system largely adheres to these trends. The inner two planets have very similar sizes, while the outer planet is nearly twice as large as the inner two. This type of configuration is well within the scatter of Figure 2 from W18. The spacing between the three planets in log semi-major axis is indeed similar, about 0.307 dex and 0.358 dex (the planets are also apparently not in resonance, with period ratios of $P_c/P_b$ = 2.8618 and $P_d/P_c$ = 3.4506). Using the mass-radius relations of \citet{Wolfgang:etal:2016} and \citet{Chen:Kipping:2017} to predict planet masses based on the radii we found $M_b \sim$ 2--5~$M_\oplus$, $M_c \sim$ 2--5~$M_\oplus$, and $M_d \sim$ 4--13~$M_\oplus$. From these predicted planet masses, the stellar mass, and orbital period ratios, we calculated that the planets in the \thestar\ system are each separated by about 30--35 mutual Hill radii, a separation larger than $\sim$80\% of adjacent pairs in three-planet systems. Observational biases are also important to consider, in that more compact systems are more likely to present multiple transiting planets. There may also be additional planets in the system that are non-transiting or below the sensitivity limits of the \ktwo\ photometry. We also note that the W18 sample does not include stars with spectral types later than K3, though we do not expect this to dramatically change any of the conclusions presented here.

\thestar\ is relatively bright and thus amenable to follow-up observations to characterize the planets in more detail. There are presently 18 (30) stars brighter than $V = 11$~mag ($K_s = 9$~mag) that host at least one known transiting planet smaller than 3~R$_\oplus$, and only 9 (13) of which host multiple transiting planets \citep{Akeson:etal:2013}.\footnote{\url{https://exoplanetarchive.ipac.caltech.edu}} Most of the bright multi-planet systems have been found with \ktwo, but this will soon change with \tess. Based on the planetary radii and our current understanding of the exoplanet mass-radius relation, the inner two planets are likely to be rocky while the outer planet is likely to have a substantial volatile envelope \citep[e.g.][]{Fulton:etal:2017}. From the planet mass estimates above, we calculated predicted radial velocity semi-amplitudes of $\sim$1.1--2.7, 0.8--1.9, and 1--3.3~\ms\ for planets b, c, and d, respectively. These amplitudes are at the limit of detectability for current instruments. 

\begin{figure}
    \centering
    \includegraphics[width=\linewidth]{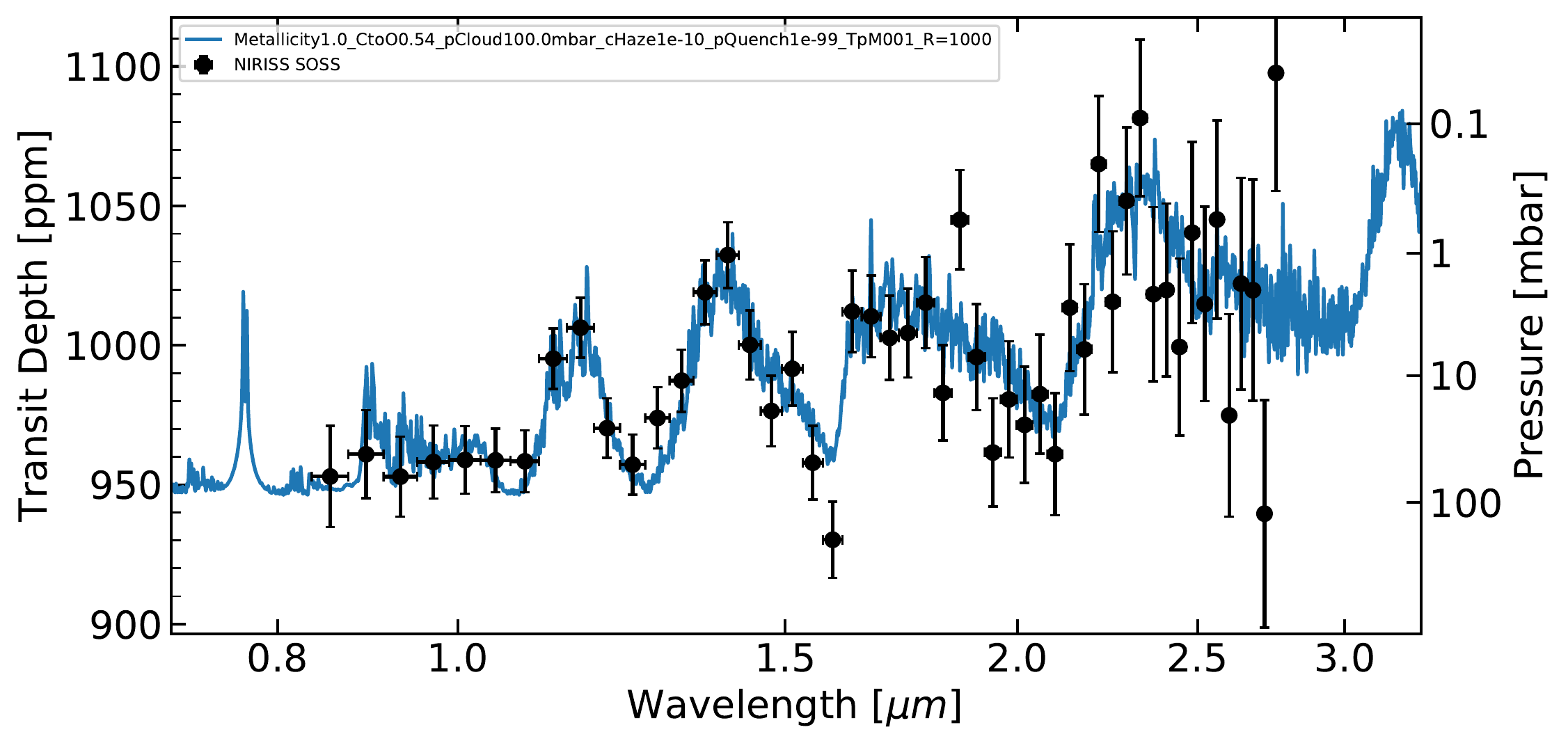}
    \caption{Model transmission spectra and simulated observations of the sub-Neptune K2-233 d, binned over 40 resolution elements resulting in $\lambda/\Delta \lambda =$ 150--250. Assuming a planetary mass of 8 $M_\oplus$ and a single transit observation by \textit{JWST/NIRISS}, water absorption is detectable at high significance for a H$_2$-dominated scenario with clouds below the 100 mbar level. Models were generated as described in \citet{Benneke:Seager:2012, Benneke:2015}. The observational uncertainties are 110\% of the photon-noise limit accounting for the exact throughput, duty-cycle, and dispersion of the instruments.}
    \label{fig:jwst}
\end{figure}

The apparent brightness and the relative small radius of the host star, \thestar, make all three planets potential targets for spectroscopic characterizations of their atmospheres. Depending on the surface gravity and hydrogen fraction of the atmosphere, the atmospheres of all three planets may be readily detectable using a single \textit{JWST} visit. We estimate transit depth variation of the order of 10--100~ppm for atmospheres dominated by ices or hydrogen/helium, respectively. Figure~\ref{fig:jwst} shows a simulated \textit{JWST/NIRISS} transmission spectrum for the sub-Neptune K2-233 d assuming a planetary mass of 8 $M_\oplus$ and hydrogen-dominated atmosphere with clouds below the 100 mbar level. Transit depth variations as a function of wavelength, predominantly due to water vapor absorption, can readily be detected.

An interesting question worthy of further exploration is to what degree do the properties and configurations of exoplanetary systems vary in time? \citet{Berger:etal:2018} recently showed that larger sub-Neptunes preferentially orbit stars younger than the Hyades age ($\sim$625--850 Myr). Such a trend might be the result of photo-evaporation, core cooling, or both. While \thestar\ is merely a single planetary system, it joins a growing sample of young exoplanet hosts from which temporal trends in planet properties can be investigated.

\begin{figure}
    \centering
    \includegraphics[width=0.45\linewidth]{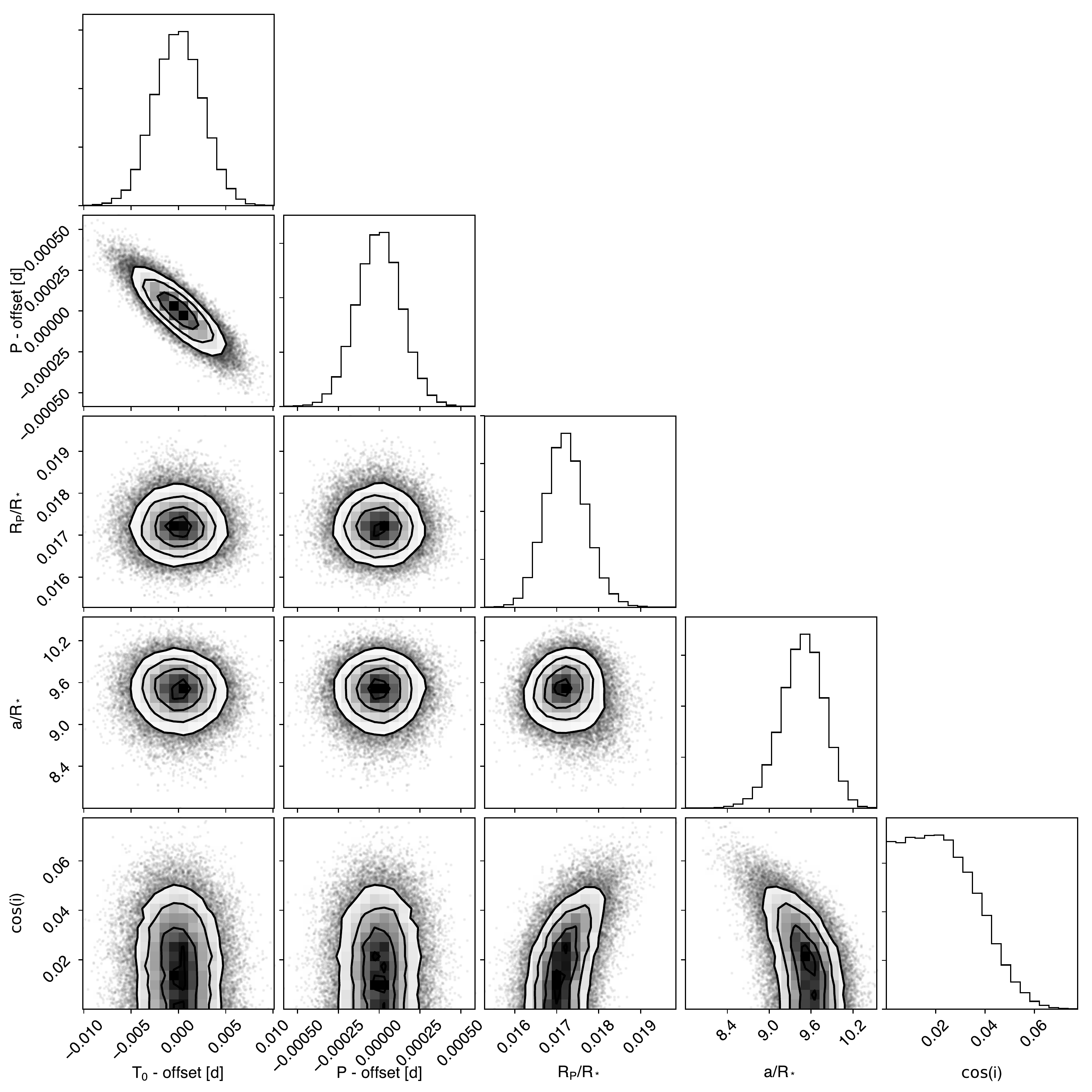}
    \includegraphics[width=0.45\linewidth]{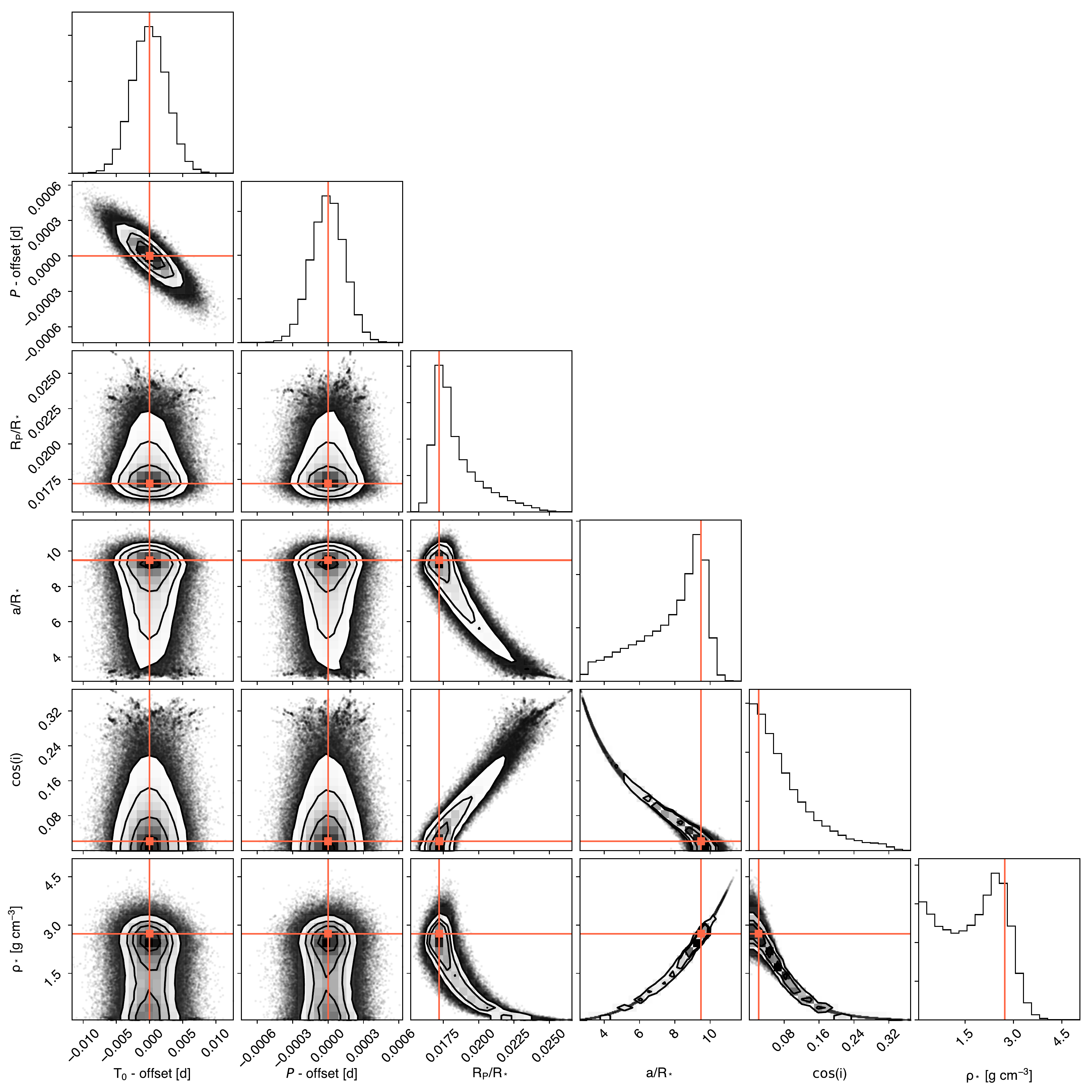}
    \caption{Corner plots from the planet b MCMC posterior samples in the case of a circular orbit with a Gaussian prior on mean stellar density (left) and with no direct prior on mean stellar density (right). At right, the positions of the red squares indicate the median values obtained (or assumed) from the fit with the imposed prior on mean stellar density.}
    \label{fig:planetb_corner}
\end{figure}

\begin{figure}
    \centering
    \includegraphics[width=0.45\linewidth]{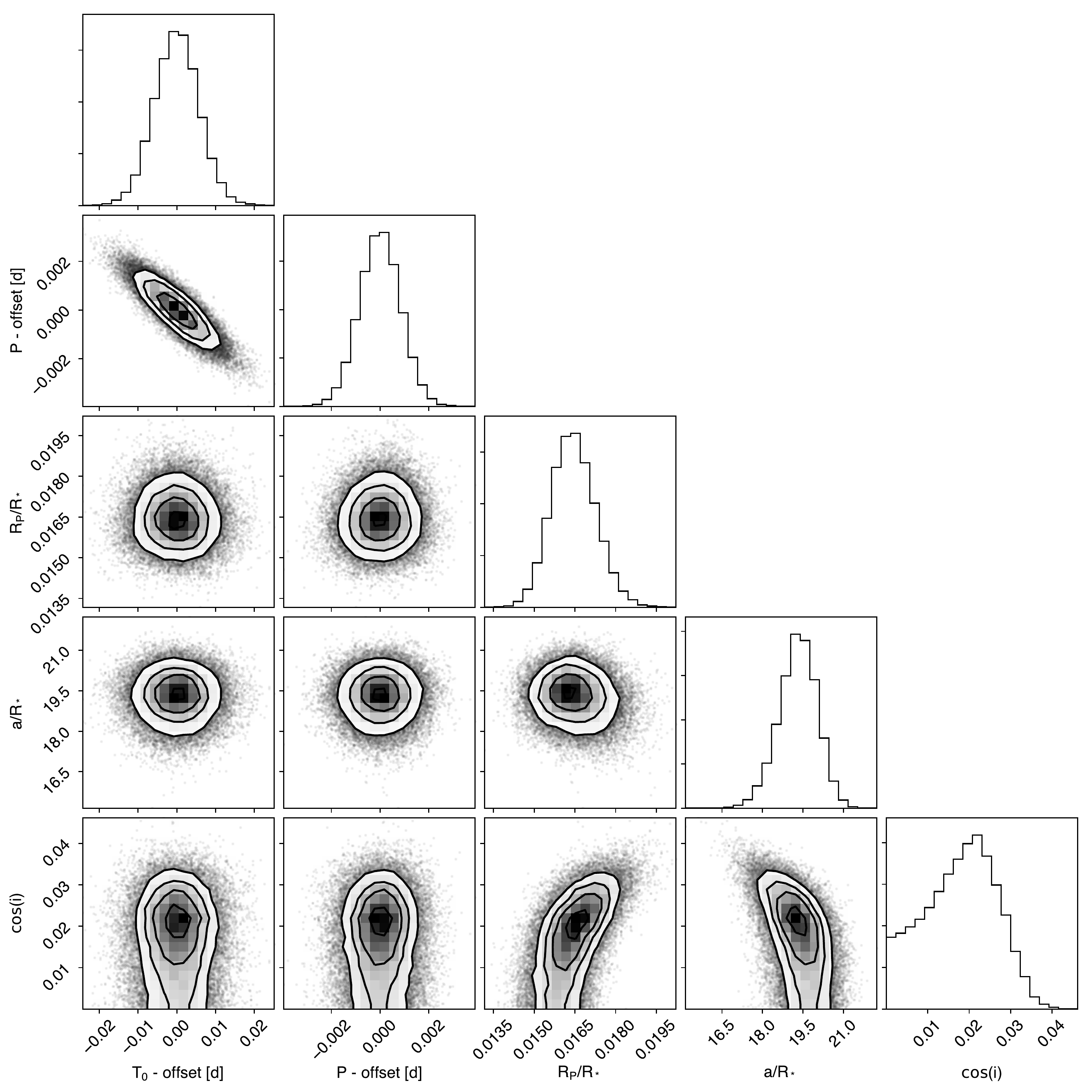}
    \includegraphics[width=0.45\linewidth]{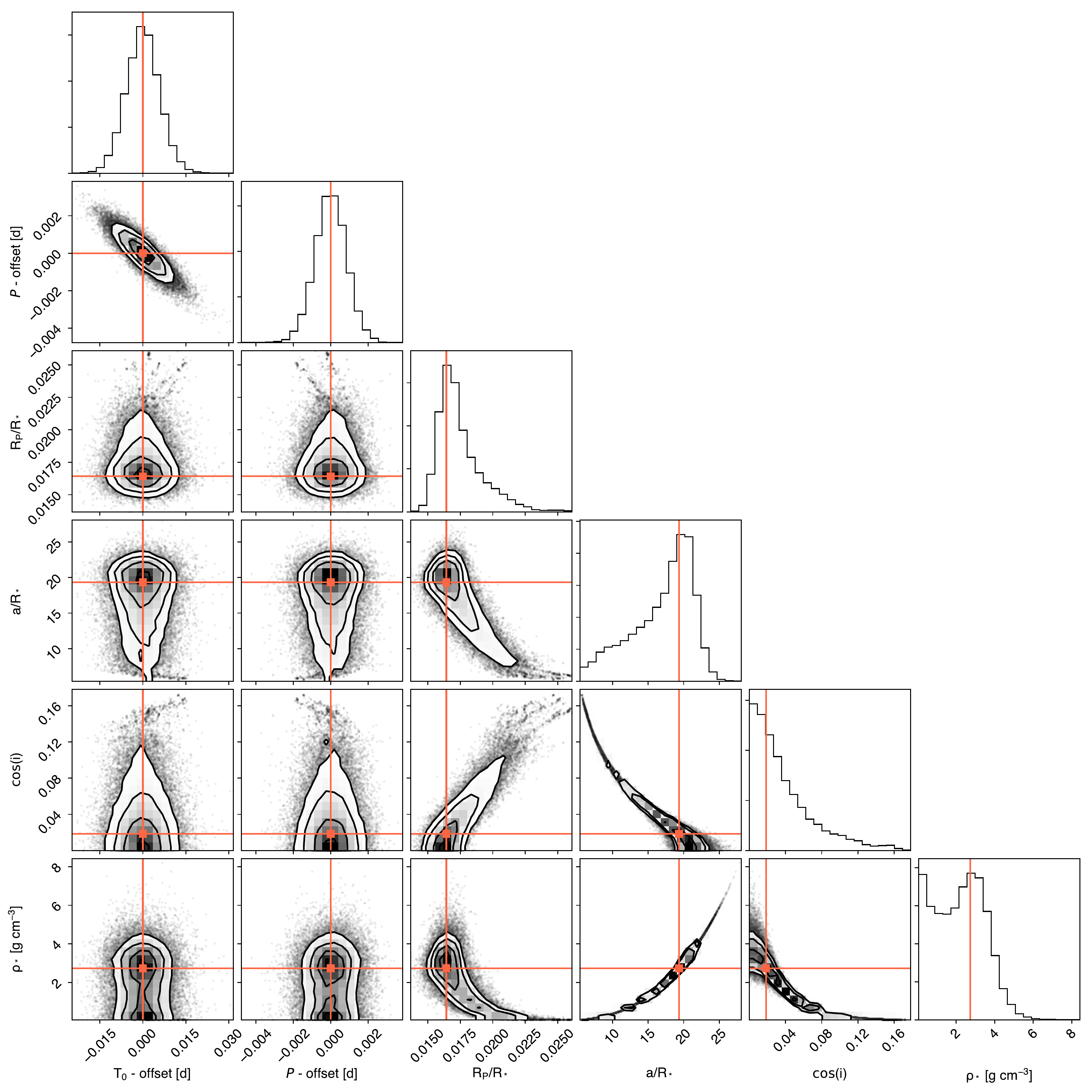}
    \caption{Same as Figure~\ref{fig:planetb_corner} for planet c.}
    \label{fig:planetc_corner}
\end{figure}

\begin{figure}
    \centering
    \includegraphics[width=0.45\linewidth]{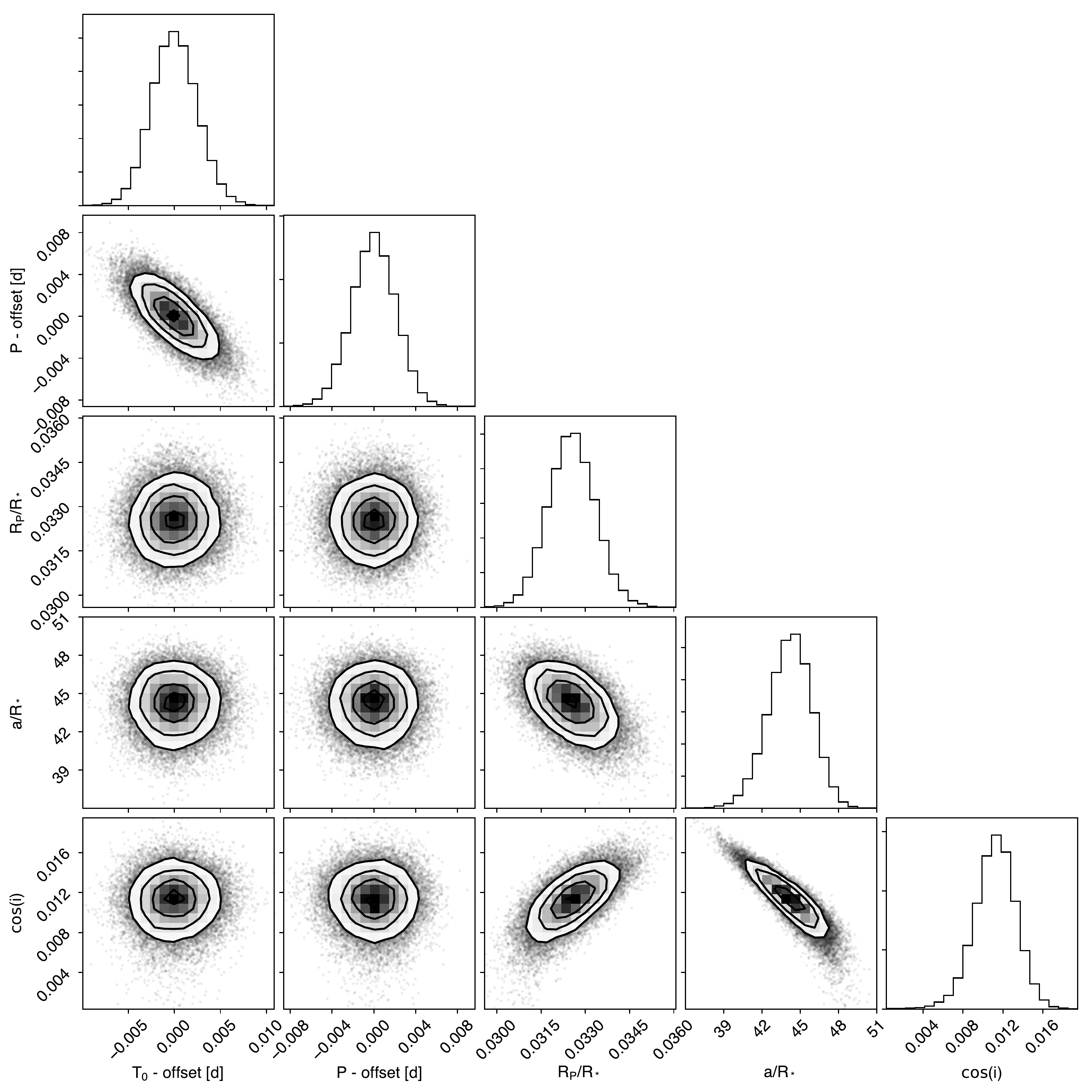}
    \includegraphics[width=0.45\linewidth]{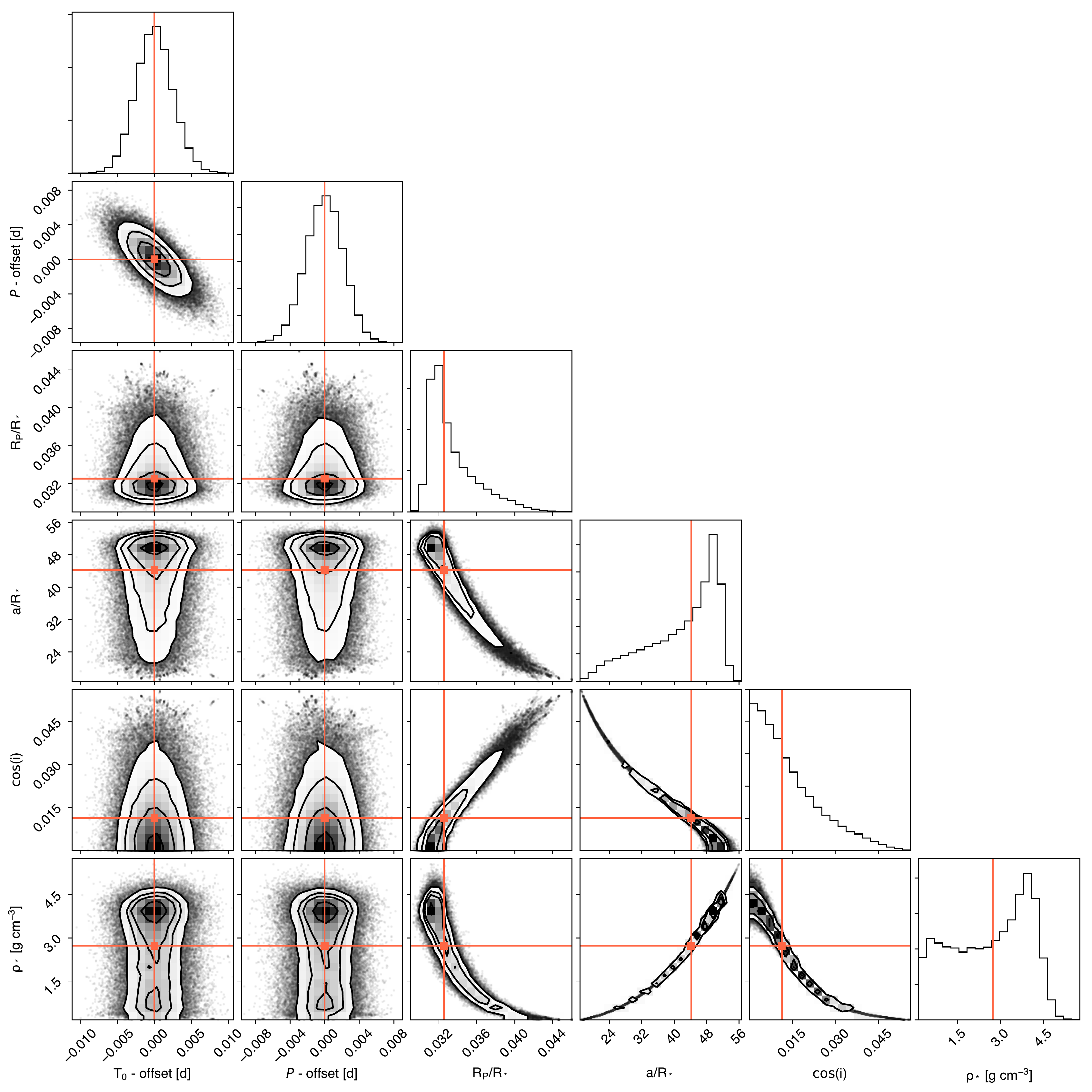}
    \caption{Same as Figure~\ref{fig:planetb_corner} for planet d.}
    \label{fig:planetd_corner}
\end{figure}

\appendix 
As mentioned in \S~\ref{sec:observations}, we performed additional transit fits for each planet with no direct prior on the mean stellar density (requiring only $a/R_*>1$ and assuming the same uniform prior on orbital period). We present the results of these fits in Table~\ref{table:fit2} and illustrate the parameter covariances from both fits in Figures~\ref{fig:planetb_corner},~\ref{fig:planetc_corner}, and \ref{fig:planetd_corner}.

\begin{deluxetable*}{lccc}
\tablecaption{Planet fit parameters with no direct prior on mean stellar density \label{table:fit2}}
\tablecolumns{4}
\tablewidth{-0pt}
\tabletypesize{\footnotesize}
\tablehead{
        \colhead{Parameter} &
        \colhead{Planet b} & 
        \colhead{Planet c} & 
        \colhead{Planet d} \\
        }
\startdata
\textit{Directly fitted parameters} \\
Time of mid-transit, $T_0$ (BJD-2450000) & 7991.6911$^{+0.0026}_{-0.0026}$ & 7996.3522$^{+0.0057}_{-0.0057}$ & 8005.5801$^{+0.0025}_{-0.0025}$ \\
Orbital period, $P$ (days) & 2.46746$^{+0.00014}_{-0.00014}$ & 7.06142$^{+0.00083}_{-0.00083}$ & 24.3662$^{+0.0021}_{-0.0022}$ \\
Radius ratio, $R_P/R_*$ & 0.01789$^{+0.00224}_{-0.00087}$ & 0.0170$^{+0.0021}_{-0.0010}$ & 0.0325$^{+0.0035}_{-0.0013}$ \\
Scaled semi-major axis, $a/R_*$ & 8.3$^{+1.2}_{-2.8}$ & 18.1$^{+2.9}_{-6.1}$ & 44.7$^{+5.6}_{-13.2}$ \\
Cosine of inclination, $\cos{i}$ & 0.061$^{+0.088}_{-0.044}$ & 0.027$^{+0.041}_{-0.020}$ & 0.0108$^{+0.0143}_{-0.0077}$ \\
\\
\textit{Derived parameters} \\
Planet radius, $R_P$ (R$_\oplus$) & 1.462$^{+0.180}_{-0.092}$ & 1.39$^{+0.17}_{-0.10}$ & 2.66$^{+0.28}_{-0.15}$ \\
Inclination, $i$ (deg) & 86.5$^{+2.5}_{-5.1}$ & 88.4$^{+1.1}_{-2.4}$ & 89.38$^{+0.44}_{-0.82}$ \\
Impact parameter, $b$ & 0.51$^{+0.32}_{-0.34}$ & 0.50$^{+0.32}_{-0.34}$ & 0.48$^{+0.31}_{-0.33}$ \\
Total duration, $T_{14}$ (hrs) & 2.01$^{+0.10}_{-0.09}$ & 2.63$^{+0.19}_{-0.17}$ & 3.82$^{+0.16}_{-0.11}$ \\
Full duration, $T_{23}$ (hrs) & 1.873$^{+0.092}_{-0.118}$ & 2.45$^{+0.18}_{-0.20}$ & 3.45$^{+0.12}_{-0.23}$ \\
Semi-major axis, $a$ (AU) & 0.03317$^{+0.00044}_{-0.00045}$ & 0.06687$^{+0.00088}_{-0.00090}$ & 0.1527$^{+0.0020}_{-0.0021}$ \\
Insolation flux, $S$ (S$_\oplus$) & 273$^{+30}_{-30}$ & 67.1$^{+7.4}_{-7.3}$ & 12.9$^{+1.4}_{-1.4}$ \\
Equilibrium temperature, $T_\mathrm{eq}$ (K)$^{a}$ & 1110$^{+248}_{-77}$ & 753$^{+172}_{-57}$ & 480$^{+91}_{-29}$ \\
Mean stellar density, $\rho_*$ (g cm$^{-3}$) & 1.81$^{+0.90}_{-1.27}$ & 2.3$^{+1.3}_{-1.6}$ & 2.9$^{+1.2}_{-1.9}$ \\
\enddata
\begin{tablenotes}
\item In this fit a circular orbit was assumed ($e$=0), with no direct prior on the mean stellar density (only the requirement that $a/R_*>1$). Reported values and 1$\sigma$ errors are the 50, 15.87, and 84.13 percentile levels from the MCMC chain. The fits presented here assumed no contaminating flux and quadratic limb darkening coefficients $u_1$ = 0.587, $u_2$ = 0.136. Derived parameters assume all three planets are orbiting the target star, and that the target star is single. 
\item (a) Assuming an albedo of 0.3. 
\end{tablenotes}
\end{deluxetable*}

\acknowledgments \copyright~2018. All rights reserved. This research was carried out at the Jet Propulsion Laboratory, California Institute of Technology, under a contract with the National Aeronautics and Space Administration. We thank Laura Kreidberg, Lauren Weiss, Dan Foreman-Mackey, and John Livingston for helpful discussions and the anonymous referee for comments which improved this manuscript. TJD and EEM gratefully acknowledge support from the Jet Propulsion Laboratory Exoplanetary Science Initiative. EEM acknowledges support from the NASA NExSS program. This Letter includes data collected by the Kepler mission. Funding for the Kepler mission is provided by the NASA Science Mission directorate. A portion of this  work was supported by a NASA Keck PI Data Award, administered by the NASA Exoplanet Science Institute. Data presented herein were obtained at the W. M. Keck Observatory from telescope time allocated to the National Aeronautics and Space Administration through the agency's scientific partnership with the California Institute of Technology and the University of California. The Observatory was made possible by the generous financial support of the W. M. Keck Foundation. The authors wish to recognize and acknowledge the very significant cultural role and reverence that the summit of Maunakea has always had within the indigenous Hawaiian community.  We are most fortunate to have the opportunity to conduct observations from this mountain.


\begin{thebibliography}{}
\expandafter\ifx\csname natexlab\endcsname\relax\def\natexlab#1{#1}\fi
\providecommand{\url}[1]{\href{#1}{#1}}
\providecommand{\dodoi}[1]{doi:~\href{http://doi.org/#1}{\nolinkurl{#1}}}
\providecommand{\doeprint}[1]{\href{http://ascl.net/#1}{\nolinkurl{http://ascl.net/#1}}}
\providecommand{\doarXiv}[1]{\href{https://arxiv.org/abs/#1}{\nolinkurl{https://arxiv.org/abs/#1}}}

\bibitem[{{Akeson} {et~al.}(2013){Akeson}, {Chen}, {Ciardi}, {Crane}, {Good},
  {Harbut}, {Jackson}, {Kane}, {Laity}, {Leifer}, {Lynn}, {McElroy}, {Papin},
  {Plavchan}, {Ram{\'{\i}}rez}, {Rey}, {von Braun}, {Wittman}, {Abajian},
  {Ali}, {Beichman}, {Beekley}, {Berriman}, {Berukoff}, {Bryden}, {Chan},
  {Groom}, {Lau}, {Payne}, {Regelson}, {Saucedo}, {Schmitz}, {Stauffer},
  {Wyatt}, \& {Zhang}}]{Akeson:etal:2013}
{Akeson}, R.~L., {Chen}, X., {Ciardi}, D., {et~al.} 2013, \pasp, 125, 989,
  \dodoi{10.1086/672273}

\bibitem[{{Angus} {et~al.}(2015){Angus}, {Aigrain}, {Foreman-Mackey}, \&
  {McQuillan}}]{Angus:etal:2015}
{Angus}, R., {Aigrain}, S., {Foreman-Mackey}, D., \& {McQuillan}, A. 2015,
  \mnras, 450, 1787, \dodoi{10.1093/mnras/stv423}

\bibitem[{Barentsen(2017)}]{geert_barentsen_2017_344973}
Barentsen, G. 2017, KeplerGO/kadenza: v2.0.2, \dodoi{10.5281/zenodo.344973}.
\newblock \url{https://doi.org/10.5281/zenodo.344973}

\bibitem[{{Barnes}(2007)}]{Barnes:2007}
{Barnes}, S.~A. 2007, \apj, 669, 1167, \dodoi{10.1086/519295}

\bibitem[{{Barrag{\'a}n} {et~al.}(2018){Barrag{\'a}n}, {Gandolfi}, {Smith},
  {Deeg}, {Fridlund}, {Persson}, {Donati}, {Endl}, {Csizmadia}, {Grziwa},
  {Nespral}, {Hatzes}, {Cochran}, {Fossati}, {Brems}, {Cabrera}, {Cusano},
  {Eigm{\"u}ller}, {Eiroa}, {Erikson}, {Guenther}, {Korth}, {Lorenzo-Oliveira},
  {Mancini}, {P{\"a}tzold}, {Prieto-Arranz}, {Rauer}, {Rebollido}, {Saario}, \&
  {Zakhozhay}}]{Barragan:etal:2018}
{Barrag{\'a}n}, O., {Gandolfi}, D., {Smith}, A.~M.~S., {et~al.} 2018, \mnras,
  475, 1765, \dodoi{10.1093/mnras/stx3207}

\bibitem[{{Benneke}(2015)}]{Benneke:2015}
{Benneke}, B. 2015, ArXiv e-prints.
\newblock \doarXiv{1504.07655}

\bibitem[{{Benneke} \& {Seager}(2012)}]{Benneke:Seager:2012}
{Benneke}, B., \& {Seager}, S. 2012, \apj, 753, 100,
  \dodoi{10.1088/0004-637X/753/2/100}

\bibitem[{{Berger} {et~al.}(2018){Berger}, {Howard}, \&
  {Boesgaard}}]{Berger:etal:2018}
{Berger}, T.~A., {Howard}, A.~W., \& {Boesgaard}, A.~M. 2018, ArXiv e-prints.
\newblock \doarXiv{1802.09529}

\bibitem[{{Biddle} {et~al.}(2018){Biddle}, {Johns-Krull}, {Llama}, {Prato}, \&
  {Skiff}}]{Biddle:etal:2018}
{Biddle}, L.~I., {Johns-Krull}, C.~M., {Llama}, J., {Prato}, L., \& {Skiff},
  B.~A. 2018, \apjl, 853, L34, \dodoi{10.3847/2041-8213/aaa897}

\bibitem[{{Chen} \& {Kipping}(2017)}]{Chen:Kipping:2017}
{Chen}, J., \& {Kipping}, D. 2017, \apj, 834, 17,
  \dodoi{10.3847/1538-4357/834/1/17}

\bibitem[{{Christiansen} {et~al.}(2018){Christiansen}, {Crossfield},
  {Barentsen}, {Lintott}, {Barclay}, {Simmons}, {Petigura}, {Schlieder},
  {Dressing}, {Vanderburg}, {Allen}, {McMaster}, {Miller}, {Veldthuis},
  {Allen}, {Wolfenbarger}, {Cox}, {Zemiro}, {Howard}, {Livingston}, {Sinukoff},
  {Catron}, {Grey}, {Kusch}, {Terentev}, {Vales}, \&
  {Kristiansen}}]{Christiansen:2018}
{Christiansen}, J.~L., {Crossfield}, I.~J.~M., {Barentsen}, G., {et~al.} 2018,
  \aj, 155, 57, \dodoi{10.3847/1538-3881/aa9be0}

\bibitem[{{Chubak} {et~al.}(2012){Chubak}, {Marcy}, {Fischer}, {Howard},
  {Isaacson}, {Johnson}, \& {Wright}}]{Chubak:etal:2012}
{Chubak}, C., {Marcy}, G., {Fischer}, D.~A., {et~al.} 2012, ArXiv e-prints.
\newblock \doarXiv{1207.6212}

\bibitem[{{Ciardi} {et~al.}(2015){Ciardi}, {Beichman}, {Horch}, \&
  {Howell}}]{Ciardi:etal:2015}
{Ciardi}, D.~R., {Beichman}, C.~A., {Horch}, E.~P., \& {Howell}, S.~B. 2015,
  \apj, 805, 16, \dodoi{10.1088/0004-637X/805/1/16}

\bibitem[{{Ciardi} {et~al.}(2013){Ciardi}, {Fabrycky}, {Ford}, {Gautier},
  {Howell}, {Lissauer}, {Ragozzine}, \& {Rowe}}]{Ciardi:etal:2013}
{Ciardi}, D.~R., {Fabrycky}, D.~C., {Ford}, E.~B., {et~al.} 2013, \apj, 763,
  41, \dodoi{10.1088/0004-637X/763/1/41}

\bibitem[{{Claret} \& {Bloemen}(2011)}]{Claret:Bloemen:2011}
{Claret}, A., \& {Bloemen}, S. 2011, \aap, 529, A75,
  \dodoi{10.1051/0004-6361/201116451}

\bibitem[{{Coelho} {et~al.}(2005){Coelho}, {Barbuy}, {Mel{\'e}ndez},
  {Schiavon}, \& {Castilho}}]{Coelho:etal:2005}
{Coelho}, P., {Barbuy}, B., {Mel{\'e}ndez}, J., {Schiavon}, R.~P., \&
  {Castilho}, B.~V. 2005, \aap, 443, 735, \dodoi{10.1051/0004-6361:20053511}

\bibitem[{{Crossfield} {et~al.}(2016){Crossfield}, {Ciardi}, {Petigura},
  {Sinukoff}, {Schlieder}, {Howard}, {Beichman}, {Isaacson}, {Dressing},
  {Christiansen}, {Fulton}, {L{\'e}pine}, {Weiss}, {Hirsch}, {Livingston},
  {Baranec}, {Law}, {Riddle}, {Ziegler}, {Howell}, {Horch}, {Everett}, {Teske},
  {Martinez}, {Obermeier}, {Benneke}, {Scott}, {Deacon}, {Aller}, {Hansen},
  {Mancini}, {Ciceri}, {Brahm}, {Jord{\'a}n}, {Knutson}, {Henning}, {Bonnefoy},
  {Liu}, {Crepp}, {Lothringer}, {Hinz}, {Bailey}, {Skemer}, \&
  {Defrere}}]{Crossfield:2016}
{Crossfield}, I.~J.~M., {Ciardi}, D.~R., {Petigura}, E.~A., {et~al.} 2016,
  \apjs, 226, 7, \dodoi{10.3847/0067-0049/226/1/7}

\bibitem[{{Crossfield} {et~al.}(2017){Crossfield}, {Ciardi}, {Isaacson},
  {Howard}, {Petigura}, {Weiss}, {Fulton}, {Sinukoff}, {Schlieder}, {Mawet},
  {Ruane}, {de Pater}, {de Kleer}, {Davies}, {Christiansen}, {Dressing},
  {Hirsch}, {Benneke}, {Crepp}, {Kosiarek}, {Livingston}, {Gonzales},
  {Beichman}, \& {Knutson}}]{Crossfield:etal:2017}
{Crossfield}, I.~J.~M., {Ciardi}, D.~R., {Isaacson}, H., {et~al.} 2017, \aj,
  153, 255, \dodoi{10.3847/1538-3881/aa6e01}

\bibitem[{{Dai} {et~al.}(2017){Dai}, {Winn}, {Gandolfi}, {Wang}, {Teske},
  {Burt}, {Albrecht}, {Barrag{\'a}n}, {Cochran}, {Endl}, {Fridlund}, {Hatzes},
  {Hirano}, {Hirsch}, {Johnson}, {Justesen}, {Livingston}, {Persson},
  {Prieto-Arranz}, {Vanderburg}, {Alonso}, {Antoniciello}, {Arriagada},
  {Butler}, {Cabrera}, {Crane}, {Cusano}, {Csizmadia}, {Deeg}, {Dieterich},
  {Eigm{\"u}ller}, {Erikson}, {Everett}, {Fukui}, {Grziwa}, {Guenther},
  {Henry}, {Howell}, {Johnson}, {Korth}, {Kuzuhara}, {Narita}, {Nespral},
  {Nowak}, {Palle}, {P{\"a}tzold}, {Rauer}, {Monta{\~n}{\'e}s
  Rodr{\'{\i}}guez}, {Shectman}, {Smith}, {Thompson}, {Van Eylen},
  {Williamson}, \& {Wittenmyer}}]{Dai:etal:2017}
{Dai}, F., {Winn}, J.~N., {Gandolfi}, D., {et~al.} 2017, \aj, 154, 226,
  \dodoi{10.3847/1538-3881/aa9065}

\bibitem[{{Fabrycky} {et~al.}(2014){Fabrycky}, {Lissauer}, {Ragozzine}, {Rowe},
  {Steffen}, {Agol}, {Barclay}, {Batalha}, {Borucki}, {Ciardi}, {Ford},
  {Gautier}, {Geary}, {Holman}, {Jenkins}, {Li}, {Morehead}, {Morris},
  {Shporer}, {Smith}, {Still}, \& {Van Cleve}}]{Fabrycky:etal:2014}
{Fabrycky}, D.~C., {Lissauer}, J.~J., {Ragozzine}, D., {et~al.} 2014, \apj,
  790, 146, \dodoi{10.1088/0004-637X/790/2/146}

\bibitem[{{Foreman-Mackey} {et~al.}(2013){Foreman-Mackey}, {Hogg}, {Lang}, \&
  {Goodman}}]{Foreman-Mackey:etal:2013}
{Foreman-Mackey}, D., {Hogg}, D.~W., {Lang}, D., \& {Goodman}, J. 2013, \pasp,
  125, 306, \dodoi{10.1086/670067}

\bibitem[{{Fulton} {et~al.}(2017){Fulton}, {Petigura}, {Howard}, {Isaacson},
  {Marcy}, {Cargile}, {Hebb}, {Weiss}, {Johnson}, {Morton}, {Sinukoff},
  {Crossfield}, \& {Hirsch}}]{Fulton:etal:2017}
{Fulton}, B.~J., {Petigura}, E.~A., {Howard}, A.~W., {et~al.} 2017, \aj, 154,
  109, \dodoi{10.3847/1538-3881/aa80eb}

\bibitem[{{Gagn{\'e}} {et~al.}(2018){Gagn{\'e}}, {Mamajek}, {Malo}, {Riedel},
  {Rodriguez}, {Lafreni{\`e}re}, {Faherty}, {Roy-Loubier}, {Pueyo}, {Robin}, \&
  {Doyon}}]{Gagne:etal:2018}
{Gagn{\'e}}, J., {Mamajek}, E.~E., {Malo}, L., {et~al.} 2018, ArXiv e-prints.
\newblock \doarXiv{1801.09051}

\bibitem[{{Gaidos} {et~al.}(2017){Gaidos}, {Mann}, {Rizzuto}, {Nofi}, {Mace},
  {Vanderburg}, {Feiden}, {Narita}, {Takeda}, {Esposito}, {De Rosa}, {Ansdell},
  {Hirano}, {Graham}, {Kraus}, \& {Jaffe}}]{Gaidos:etal:2017}
{Gaidos}, E., {Mann}, A.~W., {Rizzuto}, A., {et~al.} 2017, \mnras, 464, 850,
  \dodoi{10.1093/mnras/stw2345}

\bibitem[{{Gandolfi} {et~al.}(2017){Gandolfi}, {Barrag{\'a}n}, {Hatzes},
  {Fridlund}, {Fossati}, {Donati}, {Johnson}, {Nowak}, {Prieto-Arranz},
  {Albrecht}, {Dai}, {Deeg}, {Endl}, {Grziwa}, {Hjorth}, {Korth}, {Nespral},
  {Saario}, {Smith}, {Antoniciello}, {Alarcon}, {Bedell}, {Blay}, {Brems},
  {Cabrera}, {Csizmadia}, {Cusano}, {Cochran}, {Eigm{\"u}ller}, {Erikson},
  {Gonz{\'a}lez Hern{\'a}ndez}, {Guenther}, {Hirano}, {Su{\'a}rez
  Mascare{\~n}o}, {Narita}, {Palle}, {Parviainen}, {P{\"a}tzold}, {Persson},
  {Rauer}, {Saviane}, {Schmidtobreick}, {Van Eylen}, {Winn}, \&
  {Zakhozhay}}]{Gandolfi:etal:2017}
{Gandolfi}, D., {Barrag{\'a}n}, O., {Hatzes}, A.~P., {et~al.} 2017, \aj, 154,
  123, \dodoi{10.3847/1538-3881/aa832a}

\bibitem[{{Hadden} \& {Lithwick}(2014)}]{Hadden:Lithwick:2014}
{Hadden}, S., \& {Lithwick}, Y. 2014, \apj, 787, 80,
  \dodoi{10.1088/0004-637X/787/1/80}

\bibitem[{{Howard} {et~al.}(2010){Howard}, {Johnson}, {Marcy}, {Fischer},
  {Wright}, {Bernat}, {Henry}, {Peek}, {Isaacson}, {Apps}, {Endl}, {Cochran},
  {Valenti}, {Anderson}, \& {Piskunov}}]{Howard:etal:2010}
{Howard}, A.~W., {Johnson}, J.~A., {Marcy}, G.~W., {et~al.} 2010, \apj, 721,
  1467, \dodoi{10.1088/0004-637X/721/2/1467}

\bibitem[{{Howard} {et~al.}(2012){Howard}, {Marcy}, {Bryson}, {Jenkins},
  {Rowe}, {Batalha}, {Borucki}, {Koch}, {Dunham}, {Gautier}, {Van Cleve},
  {Cochran}, {Latham}, {Lissauer}, {Torres}, {Brown}, {Gilliland}, {Buchhave},
  {Caldwell}, {Christensen-Dalsgaard}, {Ciardi}, {Fressin}, {Haas}, {Howell},
  {Kjeldsen}, {Seager}, {Rogers}, {Sasselov}, {Steffen}, {Basri},
  {Charbonneau}, {Christiansen}, {Clarke}, {Dupree}, {Fabrycky}, {Fischer},
  {Ford}, {Fortney}, {Tarter}, {Girouard}, {Holman}, {Johnson}, {Klaus},
  {Machalek}, {Moorhead}, {Morehead}, {Ragozzine}, {Tenenbaum}, {Twicken},
  {Quinn}, {Isaacson}, {Shporer}, {Lucas}, {Walkowicz}, {Welsh}, {Boss},
  {Devore}, {Gould}, {Smith}, {Morris}, {Prsa}, {Morton}, {Still}, {Thompson},
  {Mullally}, {Endl}, \& {MacQueen}}]{Howard:etal:2012}
{Howard}, A.~W., {Marcy}, G.~W., {Bryson}, S.~T., {et~al.} 2012, \apjs, 201,
  15, \dodoi{10.1088/0067-0049/201/2/15}

\bibitem[{{Howell} {et~al.}(2014){Howell}, {Sobeck}, {Haas}, {Still},
  {Barclay}, {Mullally}, {Troeltzsch}, {Aigrain}, {Bryson}, {Caldwell},
  {Chaplin}, {Cochran}, {Huber}, {Marcy}, {Miglio}, {Najita}, {Smith},
  {Twicken}, \& {Fortney}}]{Howell:etal:2014}
{Howell}, S.~B., {Sobeck}, C., {Haas}, M., {et~al.} 2014, \pasp, 126, 398,
  \dodoi{10.1086/676406}

\bibitem[{{Huber} {et~al.}(2017){Huber}, {Zinn}, {Bojsen-Hansen},
  {Pinsonneault}, {Sahlholdt}, {Serenelli}, {Silva Aguirre}, {Stassun},
  {Stello}, {Tayar}, {Bastien}, {Bedding}, {Buchhave}, {Chaplin}, {Davies},
  {Garc{\'{\i}}a}, {Latham}, {Mathur}, {Mosser}, \& {Sharma}}]{Huber:etal:2017}
{Huber}, D., {Zinn}, J., {Bojsen-Hansen}, M., {et~al.} 2017, \apj, 844, 102,
  \dodoi{10.3847/1538-4357/aa75ca}

\bibitem[{{Isaacson} \& {Fischer}(2010)}]{Isaacson:etal:2010}
{Isaacson}, H., \& {Fischer}, D. 2010, \apj, 725, 875,
  \dodoi{10.1088/0004-637X/725/1/875}

\bibitem[{{Kolbl} {et~al.}(2015){Kolbl}, {Marcy}, {Isaacson}, \&
  {Howard}}]{Kolbl:etal:2015}
{Kolbl}, R., {Marcy}, G.~W., {Isaacson}, H., \& {Howard}, A.~W. 2015, \aj, 149,
  18, \dodoi{10.1088/0004-6256/149/1/18}

\bibitem[{{Lopez} \& {Fortney}(2013)}]{Lopez:Fortney:2013}
{Lopez}, E.~D., \& {Fortney}, J.~J. 2013, \apj, 776, 2,
  \dodoi{10.1088/0004-637X/776/1/2}

\bibitem[{{Mamajek} \& {Hillenbrand}(2008)}]{Mamajek:Hillenbrand:2008}
{Mamajek}, E.~E., \& {Hillenbrand}, L.~A. 2008, \apj, 687, 1264,
  \dodoi{10.1086/591785}

\bibitem[{{Millholland} {et~al.}(2017){Millholland}, {Wang}, \&
  {Laughlin}}]{Milholland:etal:2017}
{Millholland}, S., {Wang}, S., \& {Laughlin}, G. 2017, \apjl, 849, L33,
  \dodoi{10.3847/2041-8213/aa9714}

\bibitem[{{Morton}(2015)}]{Morton:2015}
{Morton}, T.~D. 2015, {VESPA: False positive probabilities calculator},
  Astrophysics Source Code Library.
\newblock \doeprint{1503.011}

\bibitem[{{Niraula} {et~al.}(2017){Niraula}, {Redfield}, {Dai}, {Barrag{\'a}n},
  {Gandolfi}, {Cauley}, {Hirano}, {Korth}, {Smith}, {Prieto-Arranz}, {Grziwa},
  {Fridlund}, {Persson}, {Justesen}, {Winn}, {Albrecht}, {Cochran},
  {Csizmadia}, {Duvvuri}, {Endl}, {Hatzes}, {Livingston}, {Narita}, {Nespral},
  {Nowak}, {P{\"a}tzold}, {Palle}, \& {Van Eylen}}]{Niraula:etal:2017}
{Niraula}, P., {Redfield}, S., {Dai}, F., {et~al.} 2017, \aj, 154, 266,
  \dodoi{10.3847/1538-3881/aa957c}

\bibitem[{{Parviainen}(2015)}]{Parviainen:2015}
{Parviainen}, H. 2015, \mnras, 450, 3233, \dodoi{10.1093/mnras/stv894}

\bibitem[{{Petigura}(2015)}]{Petigura:2015}
{Petigura}, E.~A. 2015, PhD thesis, University of California, Berkeley

\bibitem[{{Petigura} {et~al.}(2013{\natexlab{a}}){Petigura}, {Howard}, \&
  {Marcy}}]{Petigura:etal:2013b}
{Petigura}, E.~A., {Howard}, A.~W., \& {Marcy}, G.~W. 2013{\natexlab{a}},
  Proceedings of the National Academy of Science, 110, 19273,
  \dodoi{10.1073/pnas.1319909110}

\bibitem[{{Petigura} {et~al.}(2013{\natexlab{b}}){Petigura}, {Marcy}, \&
  {Howard}}]{Petigura:etal:2013a}
{Petigura}, E.~A., {Marcy}, G.~W., \& {Howard}, A.~W. 2013{\natexlab{b}}, \apj,
  770, 69, \dodoi{10.1088/0004-637X/770/1/69}

\bibitem[{{Rebull} {et~al.}(2017){Rebull}, {Stauffer}, {Hillenbrand}, {Cody},
  {Bouvier}, {Soderblom}, {Pinsonneault}, \& {Hebb}}]{Rebull:etal:2017}
{Rebull}, L.~M., {Stauffer}, J.~R., {Hillenbrand}, L.~A., {et~al.} 2017, \apj,
  839, 92, \dodoi{10.3847/1538-4357/aa6aa4}

\bibitem[{{Rebull} {et~al.}(2016){Rebull}, {Stauffer}, {Bouvier}, {Cody},
  {Hillenbrand}, {Soderblom}, {Valenti}, {Barrado}, {Bouy}, {Ciardi},
  {Pinsonneault}, {Stassun}, {Micela}, {Aigrain}, {Vrba}, {Somers},
  {Christiansen}, {Gillen}, \& {Collier Cameron}}]{Rebull:etal:2016}
{Rebull}, L.~M., {Stauffer}, J.~R., {Bouvier}, J., {et~al.} 2016, \aj, 152,
  113, \dodoi{10.3847/0004-6256/152/5/113}

\bibitem[{{Rizzuto} {et~al.}(2017){Rizzuto}, {Mann}, {Vanderburg}, {Kraus}, \&
  {Covey}}]{Rizzuto:etal:2017}
{Rizzuto}, A.~C., {Mann}, A.~W., {Vanderburg}, A., {Kraus}, A.~L., \& {Covey},
  K.~R. 2017, \aj, 154, 224, \dodoi{10.3847/1538-3881/aa9070}

\bibitem[{{Rodriguez} {et~al.}(2018){Rodriguez}, {Vanderburg}, {Eastman},
  {Mann}, {Crossfield}, {Ciardi}, {Latham}, \& {Quinn}}]{Rodriguez:etal:2018}
{Rodriguez}, J.~E., {Vanderburg}, A., {Eastman}, J.~D., {et~al.} 2018, \aj,
  155, 72, \dodoi{10.3847/1538-3881/aaa292}

\bibitem[{{Rodriguez} {et~al.}(2017){Rodriguez}, {Zhou}, {Vanderburg},
  {Eastman}, {Kreidberg}, {Cargile}, {Bieryla}, {Latham}, {Irwin}, {Mayo},
  {Calkins}, {Esquerdo}, \& {Mink}}]{Rodriguez:etal:2017}
{Rodriguez}, J.~E., {Zhou}, G., {Vanderburg}, A., {et~al.} 2017, \aj, 153, 256,
  \dodoi{10.3847/1538-3881/aa6dfb}

\bibitem[{{Van Eylen} \& {Albrecht}(2015)}]{vanEylen:Albrecht:2015}
{Van Eylen}, V., \& {Albrecht}, S. 2015, \apj, 808, 126,
  \dodoi{10.1088/0004-637X/808/2/126}

\bibitem[{{Vanderburg} {et~al.}(2015){Vanderburg}, {Johnson}, {Rappaport},
  {Bieryla}, {Irwin}, {Lewis}, {Kipping}, {Brown}, {Dufour}, {Ciardi}, {Angus},
  {Schaefer}, {Latham}, {Charbonneau}, {Beichman}, {Eastman}, {McCrady},
  {Wittenmyer}, \& {Wright}}]{Vanderburg:etal:2015}
{Vanderburg}, A., {Johnson}, J.~A., {Rappaport}, S., {et~al.} 2015, \nat, 526,
  546, \dodoi{10.1038/nature15527}

\bibitem[{{Vanderburg} {et~al.}(2016{\natexlab{a}}){Vanderburg}, {Becker},
  {Kristiansen}, {Bieryla}, {Duev}, {Jensen-Clem}, {Morton}, {Latham}, {Adams},
  {Baranec}, {Berlind}, {Calkins}, {Esquerdo}, {Kulkarni}, {Law}, {Riddle},
  {Salama}, \& {Schmitt}}]{Vanderburg:etal:2016b}
{Vanderburg}, A., {Becker}, J.~C., {Kristiansen}, M.~H., {et~al.}
  2016{\natexlab{a}}, \apjl, 827, L10, \dodoi{10.3847/2041-8205/827/1/L10}

\bibitem[{{Vanderburg} {et~al.}(2016{\natexlab{b}}){Vanderburg}, {Bieryla},
  {Duev}, {Jensen-Clem}, {Latham}, {Mayo}, {Baranec}, {Berlind}, {Kulkarni},
  {Law}, {Nieberding}, {Riddle}, \& {Salama}}]{Vanderburg:etal:2016c}
{Vanderburg}, A., {Bieryla}, A., {Duev}, D.~A., {et~al.} 2016{\natexlab{b}},
  \apjl, 829, L9, \dodoi{10.3847/2041-8205/829/1/L9}

\bibitem[{{Vogt} {et~al.}(1994){Vogt}, {Allen}, {Bigelow}, {Bresee}, {Brown},
  {Cantrall}, {Conrad}, {Couture}, {Delaney}, {Epps}, {Hilyard}, {Hilyard},
  {Horn}, {Jern}, {Kanto}, {Keane}, {Kibrick}, {Lewis}, {Osborne},
  {Pardeilhan}, {Pfister}, {Ricketts}, {Robinson}, {Stover}, {Tucker}, {Ward},
  \& {Wei}}]{Vogt:etal:1994}
{Vogt}, S.~S., {Allen}, S.~L., {Bigelow}, B.~C., {et~al.} 1994, in \procspie,
  Vol. 2198, Instrumentation in Astronomy VIII, ed. D.~L. {Crawford} \& E.~R.
  {Craine}, 362

\bibitem[{{Weiss} {et~al.}(2018){Weiss}, {Marcy}, {Petigura}, {Fulton},
  {Howard}, {Winn}, {Isaacson}, {Morton}, {Hirsch}, {Sinukoff}, {Cumming},
  {Hebb}, \& {Cargile}}]{Weiss:etal:2018}
{Weiss}, L.~M., {Marcy}, G.~W., {Petigura}, E.~A., {et~al.} 2018, \aj, 155, 48,
  \dodoi{10.3847/1538-3881/aa9ff6}

\bibitem[{{Wolfgang} {et~al.}(2016){Wolfgang}, {Rogers}, \&
  {Ford}}]{Wolfgang:etal:2016}
{Wolfgang}, A., {Rogers}, L.~A., \& {Ford}, E.~B. 2016, \apj, 825, 19,
  \dodoi{10.3847/0004-637X/825/1/19}

\bibitem[{{Xie} {et~al.}(2016){Xie}, {Dong}, {Zhu}, {Huber}, {Zheng}, {De Cat},
  {Fu}, {Liu}, {Luo}, {Wu}, {Zhang}, {Zhang}, {Zhou}, {Cao}, {Hou}, {Wang}, \&
  {Zhang}}]{Xie:etal:2016}
{Xie}, J.-W., {Dong}, S., {Zhu}, Z., {et~al.} 2016, Proceedings of the National
  Academy of Science, 113, 11431, \dodoi{10.1073/pnas.1604692113}

\bibitem[{{Yee} {et~al.}(2017){Yee}, {Petigura}, \& {von
  Braun}}]{Yee:etal:2017}
{Yee}, S.~W., {Petigura}, E.~A., \& {von Braun}, K. 2017, \apj, 836, 77,
  \dodoi{10.3847/1538-4357/836/1/77}

\end{thebibliography}

\end{document}